% Kodierungen zur automatischen Erstellung des Layouts:
\magnification=\magstep1
\input epsf
%\voffset= 2.0true cm
\vsize= 18 cm     % wird im Ausdruck 23.7
\hsize= 14 cm
\hfuzz=2pt
\tolerance=500
\abovedisplayskip=3 mm plus6pt minus 4pt
\belowdisplayskip=3 mm plus6pt minus 4pt
\abovedisplayshortskip=0mm plus6pt
\belowdisplayshortskip=2 mm plus4pt minus 4pt
\predisplaypenalty=0
\footline={\tenrm\ifodd\pageno\hfil\folio\else\folio\hfil\fi}
%-----------------------------------------------------------------------

\def\la{\mathrel{\hbox{\rlap{\hbox{\lower4pt\hbox{$\sim$}}}\hbox{$<$}}}}
\def\ga{\mathrel{\hbox{\rlap{\hbox{\lower4pt\hbox{$\sim$}}}\hbox{$>$}}}}

\def\arcmin{\hbox{$^\prime$}}

\def\utw{\smash{\rlap{\lower5pt\hbox{$\sim$}}}}
\def\udtw{\smash{\rlap{\lower6pt\hbox{$\approx$}}}}

\def\farcm{\hbox{$.\mkern-4mu^\prime$}}

\def\getsto{\mathrel{\hbox{\rlap{$\gets$}\hbox{\raise2pt\hbox{$\to$}}}}}
\def\lid{\mathrel{\hbox{\rlap{\hbox{\lower4pt\hbox{$=$}}}\hbox{$<$}}}}
\def\gid{\mathrel{\hbox{\rlap{\hbox{\lower4pt\hbox{$=$}}}\hbox{$>$}}}}
\def\sol{\mathrel{\hbox{\rlap{\hbox{\raise4pt\hbox{$\sim$}}}\hbox{$<$}}}
}
\def\sog{\mathrel{\hbox{\rlap{\hbox{\raise4pt\hbox{$\sim$}}}\hbox{$>$}}}
}
\def\lse{\mathrel{\hbox{\rlap{\hbox{\raise4pt\hbox{$<$}}}\hbox{$\simeq$}
}}}
\def\gse{\mathrel{\hbox{\rlap{\hbox{\raise4pt\hbox{$>$}}}\hbox{$\simeq$}
}}}
\def\grole{\mathrel{\hbox{\lower2pt\hbox{$<$}}\kern-8pt
\hbox{\raise2pt\hbox{$>$}}}}
\def\leogr{\mathrel{\hbox{\lower2pt\hbox{$>$}}\kern-8pt
\hbox{\raise2pt\hbox{$<$}}}}
\def\loa{\mathrel{\hbox{\rlap{\hbox{\lower4pt\hbox{$\approx$}}}\hbox{$<$
}}}}
\def\goa{\mathrel{\hbox{\rlap{\hbox{\lower4pt\hbox{$\approx$}}}\hbox{$>$
}}}}

%-----------------------------------------------------------------------
%
%  Fontdefinitionen
%

% vektor-fonts
%\font\halbcurs gibt es schon als \tams=cmmib10
\font\kleinhalbcurs=cmmib10 scaled 833
% petit-fonts
\font\eightrm=cmr8
\font\sixrm=cmr6
\font\eighti=cmmi8
\font\sixi=cmmi6
\skewchar\eighti='177 \skewchar\sixi='177
\font\eightsy=cmsy8
\font\sixsy=cmsy6
\skewchar\eightsy='60 \skewchar\sixsy='60
\font\eightbf=cmbx8
\font\sixbf=cmbx6
\font\eighttt=cmtt8
\hyphenchar\eighttt=-1
\font\eightsl=cmsl8
\font\eightit=cmti8

\font\bxf=cmbx10
%-------------------------------------------------------------------
% Definition der versal griechischen Buchstaben
%=======================================================================
  \mathchardef\Gamma="0100
  \mathchardef\Delta="0101
  \mathchardef\Theta="0102
  \mathchardef\Lambda="0103
  \mathchardef\Xi="0104
  \mathchardef\Pi="0105
  \mathchardef\Sigma="0106
  \mathchardef\Upsilon="0107
  \mathchardef\Phi="0108
  \mathchardef\Psi="0109
  \mathchardef\Omega="010A
%-----------------------------------------------------------------------
\def\rahmen#1{\vskip#1truecm}
% Abbildungen
\def\begfig#1cm#2\endfig{\par
\setbox1=\vbox{\rahmen{#1}#2}%
\dimen0=\ht1\advance\dimen0by\dp1\advance\dimen0by5\baselineskip
\advance\dimen0by0.4true cm
\ifdim\dimen0>\vsize\pageinsert\box1\vfill\endinsert
\else%keine seitenhohe Abbildung
\dimen0=\pagetotal\ifdim\dimen0<\pagegoal
\advance\dimen0by\ht1\advance\dimen0by\dp1\advance\dimen0by1.4true cm
\ifdim\dimen0>\vsize
\topinsert\box1\endinsert
\else\vskip1true cm\box1\vskip4true mm\fi
\else\vskip1true cm\box1\vskip4true mm\fi\fi}
%-------------------------------------------------------------------
% Abbildungslegenden
% Falls Text kleiner als eine volle Zeile, zentriert.
\def\figure#1#2{\smallskip\setbox0=\vbox{\noindent\petit{\bf Fig.\ts#1.\
}\ignorespaces #2\smallskip
\count255=0\global\advance\count255by\prevgraf}%
\ifnum\count255>1\box0\else
\centerline{\petit{\bf Fig.\ts#1.\ }\ignorespaces#2}\smallskip\fi}
%-----------------------------------------------------------------

\def\xfigure#1#2#3#4{\midinsert\noindent
    $$\epsfxsize=#4truecm\epsffile{#3}$$
    \figure{#1}{#2}\endinsert}

%-----------------------------------------------------------------
% Tabellenkoepfe

%-------------------------------------------------------------------
\def\begtab#1cm#2\endtab{\par
\ifvoid\topins\midinsert\vbox{#2\rahmen{#1}}\endinsert
\else\topinsert\vbox{#2\kern#1true cm}\endinsert\fi}
\def\rahmen#1{\vskip#1truecm}
%-----------------------------------------------------------------
\def\begpet{\vskip6pt\bgroup\petit}
\def\endpet{\vskip6pt\egroup}
% Referenzen
\def\begref{\par\bgroup\petit
\let\it=\rm\let\bf=\rm\let\sl=\rm\let\INS=N}
% Jede Referenz bleibt komplett, kein Seitenumbruch dazwischen
%\def\ref#1{\filbreak\if N\INS\let\INS=Y\vbox{\sec{References}}
%\fi\hangindent\parindent
%\hangafter=1\noindent\hbox to\parindent{#1\hfil}\ignorespaces}
%\let\endref=\endpet% Ende der Referenzen
%-------------------------------------------------------------------
%\def\vec#1{\hbox{\textfont1=\tamss\scriptfont1=\kleinhalbcurs
%\textfont0=\bxf\scriptfont0=\sevenbf
%$#1$}}
%---------------------------------------------------------------------
\def\petit{\def\rm{\fam0\eightrm}%
\textfont0=\eightrm \scriptfont0=\sixrm \scriptscriptfont0=\fiverm
 \textfont1=\eighti \scriptfont1=\sixi \scriptscriptfont1=\fivei
 \textfont2=\eightsy \scriptfont2=\sixsy \scriptscriptfont2=\fivesy
 \def\it{\fam\itfam\eightit}%
 \textfont\itfam=\eightit
 \def\sl{\fam\slfam\eightsl}%
 \textfont\slfam=\eightsl
 \def\bf{\fam\bffam\eightbf}%
 \textfont\bffam=\eightbf \scriptfont\bffam=\sixbf
 \scriptscriptfont\bffam=\fivebf
 \def\tt{\fam\ttfam\eighttt}%
 \textfont\ttfam=\eighttt
 \normalbaselineskip=9pt
 \setbox\strutbox=\hbox{\vrule height7pt depth2pt width0pt}%
 \normalbaselines\rm
\def\vec##1{\setbox0=\hbox{$##1$}\hbox{\hbox
to0pt{\copy0\hss}\kern0.45pt\box0}}}%
\let\ts=\thinspace
%-------------------------------------------------------------------
%Fonts fur die Uberschriften:
%
\font \tafontt=     cmbx10 scaled\magstep2
\font \tafonts=     cmbx7  scaled\magstep2
\font \tafontss=     cmbx5  scaled\magstep2
\font \tamt= cmmib10 scaled\magstep2
\font \tams= cmmib10 scaled\magstep1
\font \tamss= cmmib10
\font \tast= cmsy10 scaled\magstep2
\font \tass= cmsy7  scaled\magstep2
\font \tasss= cmsy5  scaled\magstep2
\font \tasyt= cmex10 scaled\magstep2
\font \tasys= cmex10 scaled\magstep1
\font \tbfontt=     cmbx10 scaled\magstep1
\font \tbfonts=     cmbx7  scaled\magstep1
\font \tbfontss=     cmbx5  scaled\magstep1
\font \tbst= cmsy10 scaled\magstep1
\font \tbss= cmsy7  scaled\magstep1
\font \tbsss= cmsy5  scaled\magstep1

%-----------------------------------------------------------------
\newbox\chsta\newbox\chstb\newbox\chstc
\def\centerpar#1{{\advance\hsize by-2\parindent
\rightskip=0pt plus 4em
\leftskip=0pt plus 4em
\parindent=0pt\setbox\chsta=\vbox{#1}%
\global\setbox\chstb=\vbox{\unvbox\chsta
\setbox\chstc=\lastbox
\line{\hfill\unhbox\chstc\unskip\unskip\unpenalty\hfill}}}%
\leftline{\kern\parindent\box\chstb}}
%---------------------------------------------------------------
 % Beginn Ueberschrift 1. Ordnung
 \def \chap#1{%\goodbreak
    \vskip24pt plus 6pt minus 4pt
    \bgroup
 \textfont0=\tafontt \scriptfont0=\tafonts \scriptscriptfont0=\tafontss
 \textfont1=\tamt \scriptfont1=\tams \scriptscriptfont1=\tamss
 \textfont2=\tast \scriptfont2=\tass \scriptscriptfont2=\tasss
 \textfont3=\tasyt \scriptfont3=\tasys \scriptscriptfont3=\tenex
     \baselineskip=18pt
     \lineskip=18pt
     \raggedright
     \pretolerance=10000
     \noindent
     \tafontt
     \ignorespaces#1\vskip7true mm plus6pt minus 4pt
     \egroup\noindent\ignorespaces}%
%------------------------------------------------------
 % Beginn Ueberschrift 2. Ordnung
 \def \sec#1{%\goodbreak
     \vskip25true pt plus4pt minus4pt
     \bgroup
 \textfont0=\tbfontt \scriptfont0=\tbfonts \scriptscriptfont0=\tbfontss
 \textfont1=\tams \scriptfont1=\tamss \scriptscriptfont1=\kleinhalbcurs
 \textfont2=\tbst \scriptfont2=\tbss \scriptscriptfont2=\tbsss
 \textfont3=\tasys \scriptfont3=\tenex \scriptscriptfont3=\tenex
     \baselineskip=16pt
     \lineskip=16pt
     \raggedright
     \pretolerance=10000
     \noindent
     \tbfontt
     \ignorespaces #1
     \vskip12true pt plus4pt minus4pt\egroup\noindent\ignorespaces}%
%------------------------------------------------------
 % Beginn Ueberschrift 3. Ordnung
 \def \subs#1{%\goodbreak
     \vskip15true pt plus 4pt minus4pt
     \bgroup
     \bxf
     \noindent
     \raggedright
     \pretolerance=10000
     \ignorespaces #1
     \vskip6true pt plus4pt minus4pt\egroup
     \noindent\ignorespaces}%
%------------------------------------------------------
 % Beginn Ueberschrift 4. Ordnung
 \def \subsubs#1{%\goodbreak
     \vskip15true pt plus 4pt minus 4pt
     \bgroup
     \bf
     \noindent
     \ignorespaces #1\unskip.\ \egroup
     \ignorespaces}
%-------------------------------------------------------------------
\def\footnoterule{\kern-3pt\hrule width 2true cm\kern2.6pt}
% Fussnoten-macros
\newcount\footcount \footcount=0
\def\advftncnt{\advance\footcount by1\global\footcount=\footcount}
% Automatisch numerierte Fussnote, Fussnotentex in petit
\def\fonote#1{\advftncnt$^{\the\footcount}$\begingroup\petit
       \def\textindent##1{\hang\noindent\hbox
       to\parindent{##1\hss}\ignorespaces}%
\vfootnote{$^{\the\footcount}$}{#1}\endgroup}
%-------------------------------------------------------------------
% Acknowledgement

%-------------------------------------------------------------------
% Satz fur bye:
\newcount\sterne
\outer\def\byebye{\bigskip\typeset
\sterne=1\ifx\speciali\undefined\else
\bigskip Special caracters created by the author
\loop\smallskip\noindent special character No\number\sterne:
\csname special\romannumeral\sterne\endcsname
\advance\sterne by 1\global\sterne=\sterne
\ifnum\sterne<11\repeat\fi
\vfill\supereject\end}
\def\typeset{\centerline{\petit This article was processed by the author
using the \TeX\ Macropackage from Springer-Verlag.}}

%%%%%%%%%%%%%%%%%%%%%%
\def\ref#1{\lbrack #1\rbrack}
\def\eck#1{\left\lbrack #1 \right\rbrack}
\def\eckk#1{\bigl[ #1 \bigr]}
\def\rund#1{\left( #1 \right)}
\def\abs#1{\left\vert #1 \right\vert}

\def\part#1#2{{\partial #1\over\partial #2}}

\def\U{{\cal U}}
\def\d{{\rm d}}

\def\eps{{\epsilon}}

\def\Real{{\rm I\mathchoice{\kern-0.70mm}{\kern-0.70mm}{\kern-0.65mm}%
  {\kern-0.50mm}R}}
  % Symbol fuer reelle Zahlen.                                 MJL
\def\C{\rm C\kern-.42em\vrule width.03em height.58em depth-.02em
       \kern.4em}
\font \bolditalics = cmmib10
\def\bx#1{\leavevmode\thinspace\hbox{\vrule\vtop{\vbox{\hrule\kern1pt
        \hbox{\vphantom{\tt/}\thinspace{\bf#1}\thinspace}}
      \kern1pt\hrule}\vrule}\thinspace}

\def \vc #1{{\textfont1=\bolditalics \hbox{$\bf#1$}}}
{\catcode`\@=11
\gdef\SchlangeUnter#1#2{\lower2pt\vbox{\baselineskip 0pt \lineskip0pt
  \ialign{$\m@th#1\hfil##\hfil$\crcr#2\crcr\sim\crcr}}}
  % kopiert von \@vereq aus dem TeXbook, Seite 360.
}

\def\ueber#1#2{{\setbox0=\hbox{$#1$}%
  \setbox1=\hbox to\wd0{\hss$\scriptscriptstyle #2$\hss}%
  \offinterlineskip
  \vbox{\box1\kern0.4mm\box0}}{}}

\def\bx#1{\leavevmode\thinspace\hbox{\vrule\vtop{\vbox{\hrule\kern1pt
        \hbox{\vphantom{\tt/}\thinspace{\bf#1}\thinspace}}
      \kern1pt\hrule}\vrule}\thinspace}

\def\SFB{{This work was supported by the "Sonderforschungsbereich 375-95 f\"ur
Astro-Teilchenphysik" der Deutschen Forschungsgemeinschaft.}}

\def\abs#1{\left\vert #1 \right\vert}

\def\nix#1{{}}

%%%%%%%%%%%%%%%%%%%%%%%
\chap{A new finite-field mass reconstruction algorithm}
\centerline{\bf Stella Seitz$^{1,2}$ \& Peter Schneider$^1$}
$^1$ Max-Planck-Institut f. Astrophysik, Postfach 1523, D-85740
Garching, Germany\hfill\break
\noindent
$^2$ Universit\"atssternwarte M\"unchen, Scheinerstr. 1, D-81679
M\"unchen, Germany

\sec{Abstract}
A new method for the reconstruction of the projected mass distribution
of clusters of galaxies from the image
distortion of background galaxies is discussed. This method is
essentially equivalent to the one we developed previously, i.e., the
noise-filtering method, but has several practical advantages:
(1) It is much easier to implement; (2) it can be easily applied to
wide-field images, since the constraints on the number of gridpoints
are much weaker than for the previous method, and (3) it can be easily
generalized to more complicated field geometries, such as that of the
Wide Field Planetary Camera 2 (WFPC2) onboard HST. We have tested the
performance of our new inversion method (for which a FORTRAN-77
implementation is available from the authors) using
simulated data, demonstrating that it fares very favourably.

\sec{1 Introduction}
The distortion of the images of background galaxies (Tyson, Valdes \&
Wenk 1990) by the tidal gravitational field of clusters of galaxies
can be used to obtain a parameter-free reconstruction of the surface
mass density of the cluster (Kaiser \& Squires 1993). Several
modifications of the original reconstruction method were proposed,
e.g., to account for distortions which are not weak (Seitz \&
Schneider 1995; Kaiser 1995), to allow an unbiased mass reconstruction
on a finite field (Schneider 1995; Kaiser et al.\ts 1995; Bartelmann
1995; Bartelmann et al.\ts 1996; Seitz \& Schneider 1996, hereafter
Paper~I; Squires \& Kaiser 1996), and to account for a broad redshift
distribution of the background galaxies (Seitz \& Schneider 1997). 
In this paper, we shall reconsider the second of the above mentioned
effects, namely mass reconstructions from data on a finite field. In
Paper~I we have derived a direct mass inversion method which is
singled out of the infinitely-many unbiased reconstructions by
identifying a component of the noise (which is due to the intrinsic
ellipticity distribution of the sources, the discreteness of galaxy
images, and observational effects) as such and filtering it out. This
noise-filter reconstruction has fared very well in numerical
simulations carried out to compare various finite-field inversions
(Paper~I; Squires \& Kaiser 1996). 

Here, we shall present a slightly revised version of the noise-filter
inversion method, which removes some of the technical drawbacks of the
original formulation. In particular, our new method can be applied to
arbitrarily-shaped data fields (which is of great interest given
the geometry of the WF chips of the WFPC2 on-board HST) and can be
used with better resolution than the previous formulation. In
addition, the numerical encoding of the new version is substantially
easier and requires much less memory. We shall formulate the inversion
problem and its solution in Sect.\ts 2, and present some practical
issues in Sect.\ts 3. Numerical tests of this method in comparison to
other reconstruction 
methods are presented in Sect.\ts 4, and we summarize in Sect.\ts 5
our findings. One application of our new method is presented in the
mass reconstruction of the cluster MS1358+62 by Hoekstra et al.\ts (1998).

Just before finalizing this manuscript, Lombardi \& Bertin (1998)
submitted a paper to the astro-ph preprint 
server. Two results of that paper are
particularly relevant for the present discussion: They have shown that
of all (direct) finite-field mass reconstructions, those with
vanishing curl in the kernel $\vc H$ -- see eq.\ts (6) below -- have
the smallest rms error; requiring that noise-free data yield an exact
mass reconstruction, they rederived the inversion method of Paper~I. 
Second, they have independently derived our new inversion method,
eq.\ts (7) below, from a variational principle.

\sec{2 Noise-filtered finite-field mass inversion}
We shall assume for simplicity that all source galaxies can be
described as being at the same redshift; this is not a necessary
assumption (see Seitz \& Schneider 1997), but simplifies the following
treatment considerably. Then,
let the mass distribution of the cluster be described by the
dimensionless surface mass density $\kappa(\vc\theta)$, and the
corresponding deflection potential be denoted by $\psi(\vc\theta)$,
such that the two-dimensional Poisson equation $\nabla^2 \psi=2\kappa$
is satisfied. The two components of the complex shear $\gamma=\gamma_1
+{\rm i}\gamma_2$ are given in terms of the deflection potential by 
$$
\gamma_1={1\over 2}\rund{\psi_{,11}-\psi_{,22}}\;,\quad
\gamma_2=\psi_{,12}\;,
\eqno (1)
$$
where indices separated by a comma denote partial
derivatives.\fonote{Note that we have changed the sign convention
compared to Paper~I.} The complex reduced shear
$$
g(\vc\theta)={\gamma(\vc\theta)\over 1-\kappa(\vc\theta)}
\eqno (2)
$$
is the expectation value of the observed image ellipticities $\eps$,
so that the observed image ellipticities provide an unbiased estimate
of the local value of $g$, as long as the cluster is non-critical. We
shall make this assumption here, although it also can be dropped (see
Seitz \& Schneider 1997). As pointed out by Schneider \& Seitz (1995),
the mass-sheet degeneracy (Gorenstein, Falco \& Shapiro 1988) allows
one to determine $(1-\kappa)$ only up to a multiplicative constant, if
no magnification information is used (Broadhurst, Taylor \& Peacock
1995; Bartelmann \& Narayan 1995). Defining 
$$
K(\vc\theta):=\ln \eckk{1-\kappa(\vc\theta)}\;,
\eqno (3)
$$
then $K$ can only be determined up to an additive constant.
Kaiser (1995) derived a relation between the gradient of $K$ and
combinations of first derivatives of $g$,
$$
\nabla K={-1\over 1-\abs{g}^2}\pmatrix{1-g_1 & -g_2\cr -g_2 &
1+g_1\cr} \pmatrix{g_{1,1}+g_{2,2} \cr g_{2,1}-g_{1,2}}
\equiv \vc u(\vc\theta)\; .
\eqno (4)
$$
The right-hand-side of this equation can be considered as an
observable, obtained from local averages of image ellipticities and by
finite differencing the resulting field $g$. 

Equation (4) can be solved (up to an additive constant) by line
integration, and several schemes for this have been proposed
(Schneider 1995; Kaiser et al.\ts 1995; Bartelmann 1995; Squires \&
Kaiser 1996). The reason why different schemes yield different results
can be seen by noting that the vector field $\vc u$ comes from (noisy)
observational estimates, and thus will in general not be a gradient
field. Therefore, the equation $\nabla K=\vc u$ has no solution in
general, since $\vc u$ has a rotational component due to observational
noise. On the other hand, if $\vc u$ is a gradient field, then all
line integration schemes are equivalent.

In Paper~I, we split the vector field into a gradient part and a
rotational part,
$$
\vc u(\vc\theta)=\nabla\tilde K(\vc\theta)+{\bf rot}\,s(\vc\theta)
\equiv \nabla\tilde K(\vc\theta)+\pmatrix{\partial s/\partial \theta_2
\cr
-\partial s/\partial \theta_1\cr }\; ,
\eqno (5)
$$
where $s(\vc\theta)$ is a scalar field. This decomposition
is not unique. However, since the rotational component is due solely
to noise, we can specify the decomposition uniquely by requiring that
the mean of ${\bf rot}\,s$ over the finite data field $\U$
vanishes, and that ${\bf rot}\,s$ vanishes if $\vc u$ is a gradient
field. These two conditions are satisfied if we set $s={\rm const}$ on
the boundary $\partial \U$ of the data field $\U$. Then, identifying
$\nabla \tilde K$ with $\nabla K$, the solution of (4) with the
rotational component removed from $\vc u$ becomes
$$
K(\vc\theta)-\bar K=\int_\U \d^2\theta'\;\vc
H(\vc\theta',\vc\theta)\cdot \vc u(\vc\theta')\; ,
\eqno (6)
$$
where $\vc H(\vc\theta',\vc\theta)$ is a vector field which can be
obtained from the Greens function of a 
Laplace equation with Neumann boundary
conditions. In Paper~I we have derived this equation and presented
explicit solutions for the cases that $\U$ is a circle or a rectangle;
in these cases, the Greens function can be obtained analytically using
geometrical methods. 

Whereas the method of Paper~I passed all numerical tests, it has a few
features which are unwanted: (1) If the geometry deviates from that of
a circle or a rectangle, the Greens function can no longer be obtained
analytically. However, a numerical determination of the Greens
function is impractical owing to its singularity. For this reason, the
mass reconstruction of the cluster Cl 0939+4713 from WFPC2 data (Seitz
et al.\ts 1996) was carried out by splitting the field into two
rectangles and combine them appropriately in the overlap region. This
is certainly not the optimal method, since each of the two individual
reconstructions made no use of the shear information outside the
respective rectangle. (2) If a quadratic field is covered by an
$N\times N$ grid of $\vc\theta$ and $\vc\theta'$ values, the necessary
memory for storing $\vc H$ consists of $2 N^4$ real numbers. Hence, if
one increases $N$ beyond $\sim 50$, the memory requirement quickly
approaches the capacity of commonly used workstations. However, due to
the singularity of the Greens function, one likes to have small grid
spacings to obtain an accurate numerical estimate of the integral (6)
-- see Squires \& Kaiser (1996) for comments on this point. (3)
Although the solution for $\vc H$ was given explicitly in Paper~I, it
is complicated and not easily encoded (though  quickly
evaluated). In order for the noise-filtering method to become a
standard and readily available tool, one would like to have an easier
method to solve for $K$.

These three points can be avoided in the following simple manner:
Taking the divergence of (5) leads to
$$
\nabla^2 K=\nabla\cdot \vc u\; .
\eqno (7a)
$$
Since $s={\rm const}$ on the boundary, ${\bf rot}\ts s$ is
perpendicular to the normal vector $\vc n$ at the boundary of $\U$, so
that
$$
\vc n\cdot \nabla K=\vc n\cdot \vc u\quad {\rm on}\;\partial\U\; .
\eqno (7b)
$$
Hence, $K$ can be obtained as the solution of the Neumann problem
given by (7a,b). There are efficient and quick methods for a
numerical solution of this problem; we have
employed a relaxation method with successive overrelaxation (see Press
et al.\ts 1992, p.857). Choosing the overrelaxation parameter as in
equation (19.5.21) of Press et al.\ts (1992), a stable solution was
found after about $20 N$ iterations on an $N\times N$ grid in
$\vc\theta$. 

The previously mentioned drawbacks of the method presented in Paper~I
are now avoided. The Neumann problem (7) can be solved for any
geometry; for example, for the WF-part of the WFPC2 one merely needs to
formulate the boundary condition (7b) at 6 sides, instead of 4 for a
rectangle. The memory requirement is reduced to a few times $N^2$ real
numbers, so that $N$ can easily be of order a few hundred. In fact,
for $N=200$, the solution of (7) takes about 2 minutes on an IBM risc
6000 processor. And finally, the numerical code for solving (7)
shrinks tremendously compared to that needed to evaluate $\vc H$.

\sec{3 Practical implementation}
In order to obtain a mass reconstruction from galaxy ellipticitities,
the following three steps are needed:

(1) The galaxy ellipticities are spatially smoothed to obtain an
unbiased estimate of the local reduced shear. If $\eps_i$ is the
complex ellipticity of the $i$-th galaxy at position $\vc\theta_i$,
and $\Delta\theta$ is a smoothing scale, we calculate $g$ at a 
position $\vc\theta$ as
$$
g(\vc\theta)={\sum_{i=1}^{N_{\rm g}} w(\abs{\vc\theta-\vc\theta_i})
\,\eps_i \over \sum_{i=1}^{N_{\rm g}}
w(\abs{\vc\theta-\vc\theta_i})}\; ,
\eqno (8)
$$
where the weight function $w$ is chosen to be 
$$
w(x)=\exp\rund{-{x^2\over \Delta\theta^2}}-\exp\rund{-q}
\eqno (9)
$$
for $x\le \sqrt{q}\Delta\theta$, and zero otherwise. This choice makes $w$
nearly Gaussian and continuous at $x=\sqrt{q}\Delta\theta$, which can be an
essential aspect when the derivatives of the components of $g$ need to
be evaluated with finite differencing and $q$ is small. In the following 
we will choose  $q=9$, where (9) becomes almost equal to the case where 
the correction term is omitted. With (8), the reduced shear $g$
can be calculated on a regular grid in $\vc\theta$. 

(2) The vector field $\vc u$ is obtained from $g$ using (4). Finite
differencing is employed, with one-sided second-order differentiation
rules taken at the boundary $\partial \U$. A further differentiation
then yields $\nabla\cdot \vc u$.

(3) The Neumann problem (7) is then solved, using the method described
above.\fonote{A Fortran 77 code of these three steps is available from
the authors by request, both for a rectangular data field, and the WFPC2
geometry.}

\sec{4 Tests and simulations}
One might wonder whether the mass reconstruction obtained with the
method described above yields smooth mass profiles. Our method requires
differentiation of (noisy) data, so it might be suspected that the
resulting mass distribution will be quite noisy compared to the
results of some of the other finite-field inversion methods. These
issues have been discussed in some detail by Squires \& Kaiser (1996);
whereas it is not a priori evident that these numerical
differentiations are unharmful to the resulting mass reconstruction,
the numerical simulations these authors have carried have shown, in
agreement with Paper~I, that
the noise-filter inversion as presented in Paper~I yields the least
noisy mass estimates of all unbiased 
{\it direct} finite-field mass reconstructions
that they have tested.\fonote{Inverse methods, such as the maximum
likelihood method (Bartelmann et al.\ts 1996) or the maximum
probability method (Squires \& Kaiser 1996) can yield slightly more
accurate mass profiles.} Since the method proposed here involves a
further differentiation of the data, we have to check whether the
noise level of the reconstruction is not increased by that.
\xfigure{1}{
Contour- and surface plot for the function $- K(\vc x) = - \ln \eck{1
-\kappa (\vc x)} $, where $\kappa$ is the two-dimensional surface mass
density of the lens. The size of the field is $7.\arcmin 5 \times
7.\arcmin 5$, and $K$ is calculated on a $50
\times 50$ grid; the spacing in the contour lines is 0.05  }
{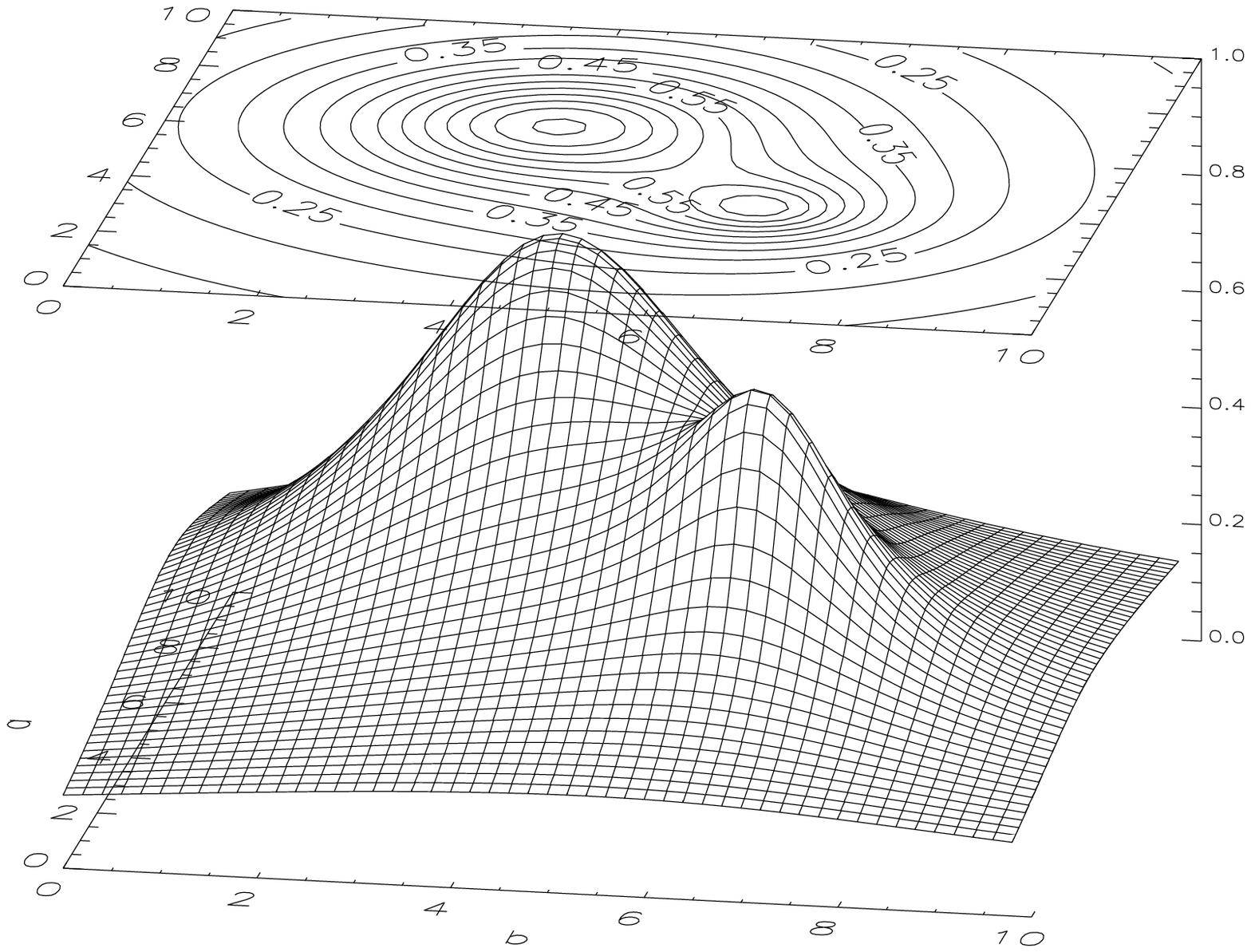 } {8}
\xfigure{2}{
Power spectra of the reconstruction error using the mass distribution
of Fig.\ts 1, a galaxy density of 50 per sq. arcminute and a width in
the ellipticty distribution of $0.2$. The solid, dotted,
dashed-dotted, dashed and long-dashed curves denote the power spectra
obtained using the methods of KS, Paper~I (NF),  Schneider (1995) and
the two versions of the newly developed noise-filtering inversion,
NF1 and NF2, respectively. The gridsize of the final reconstruction
was varied from $N=30$ to $N=80$.   }
{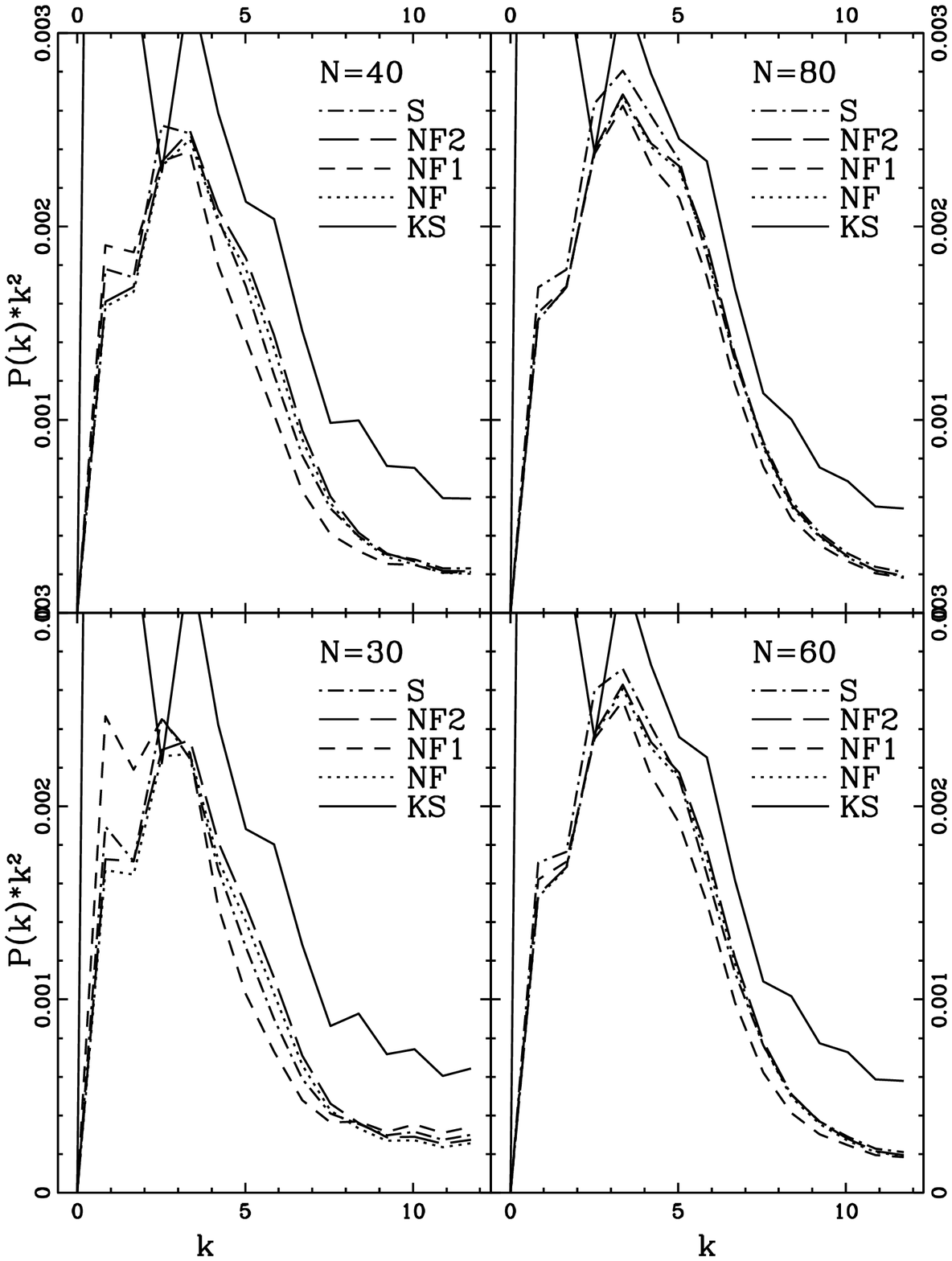 } {13}
For this purpose, we have carried out two sets of simulations. In both
cases, galaxies were distributed randomly on the data field $\U$, with
a density of 50 galaxies per square arcminute and an intrinsic ellipticity
distribution which is assumed to be a Gaussian of width $\rho =0.2$
(see Paper~I). In the first set of simulations, a mass distribution
for the cluster corresponding to the lens model B in Paper~I was
assumed (see Fig.\ts 1), and the `observed' ellipticities were
calculated from the intrinsic ellipticities and the local value of the
reduced shear caused by the lens model. In the second set of
simulations, no lens was assumed; owing to the intrinsic ellipticity
of the sources, the reduced shear as calculated from the `observed'
ellipticity does not vanish identically, and so the reconstructed mass
profile will be different from zero (this is the kind of simulations
carried out in Squires \& Kaiser 1996). The smoothing length was set
to $\Delta \theta=0\farcm 35$.

Reconstructions for the case with a lens were performed using the
following methods: the original Kaiser \& Squires (1993)
reconstruction, generalized to account for non-linear effects as
described in Paper~I; a finite-field reconstruction based on line
integration (Schneider 1995); the noise-filtering method as described
in Paper~I; and the new noise-filtering method as presented here. The
reconstructions were analyzed by Fourier-decomposition of their
difference from the input mass distribution (or, more precisely, the
input field of $K$).  From 50 different realizations of the galaxy
population, the power spectrum of this difference was obtained, using
the same procedure as in Paper~I. In Fig.\ts 2, we have plotted these
power spectra for reconstructions using 40, 80, 30 and 60 gridpoints.
Since the KS-reconstruction leads to systematic artefacts for finite fields 
with non-zero shear field outside the data region, its power spectrum (solid
line) exceeds that of all exact finite field inversion techniques (see Paper~I 
for more details). The power spectra for the reconstructions using the 
method of Schneider (1995) and the noise-filtering technique of Paper~I (NF) 
are shown as dashed-dotted and dotted curves, respectively.
The noise filtering inversion 
developed here was implented in two versions (NF1,NF2); in the first
case (short-dashed curves) (7a\&b) was solved on the same grid as for the other
inversion techniques. 
In the second case (long-dashed curves), the solution for $K$ was obtained 
on a grid two times as dense
and $K$ was estimated on the sparser grid afterwards.
\xfigure{3}{The same
power spectra as in Fig.\ts 2, but without any lens in the field,
i.e.\ts $\kappa = 0 =\gamma $  } 
{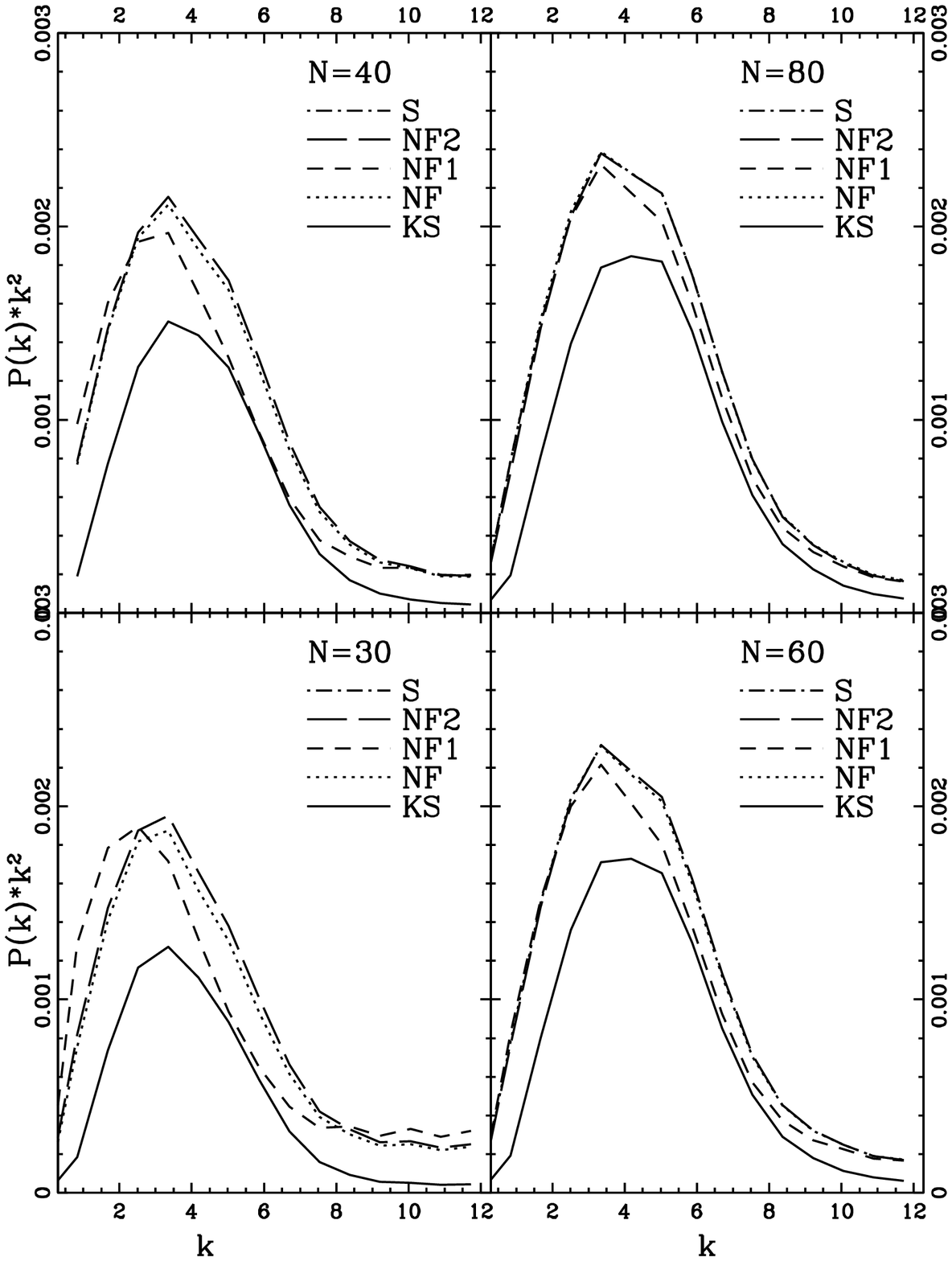 } {13}
It turns out that reconstructions with the new technique developed
here are always smoother than any of the other methods considered
here. This is because $\nabla\cdot \vc u$ is used instead of $\vc u$
itself. The operation $\nabla\cdot \vc u$ effectively yields a loss of
signal and noise on length scales of two grid points. To obtain
reconstructions with the same resolution we thus double the number of
gridpoints in NF2, calculate $\vc u$ and its divergence and $K$ on the
dense grid. Finally $K$ is calculated on the sparser grid by averaging
over 4 gridpoints. The power spectra of these NF2 reconstructions
(long dashed curves) are always very similar to the original
NF-reconstruction. Those of the NF1 reconstructions are more similar
to reconstructions on a sparser grid, where the high frequency power
is reduced due to the loss of degrees of freedom.  Since the dotted
and long-dashed curves in Fig.\ts 2 are almost the same, this
demontrates that the recovery of the signal and the sensitivity to the
noise in the NF2 and NF-method are identical.
\xfigure{4}{The power spectra of various mass inversion methods, as
explained in the main text}
{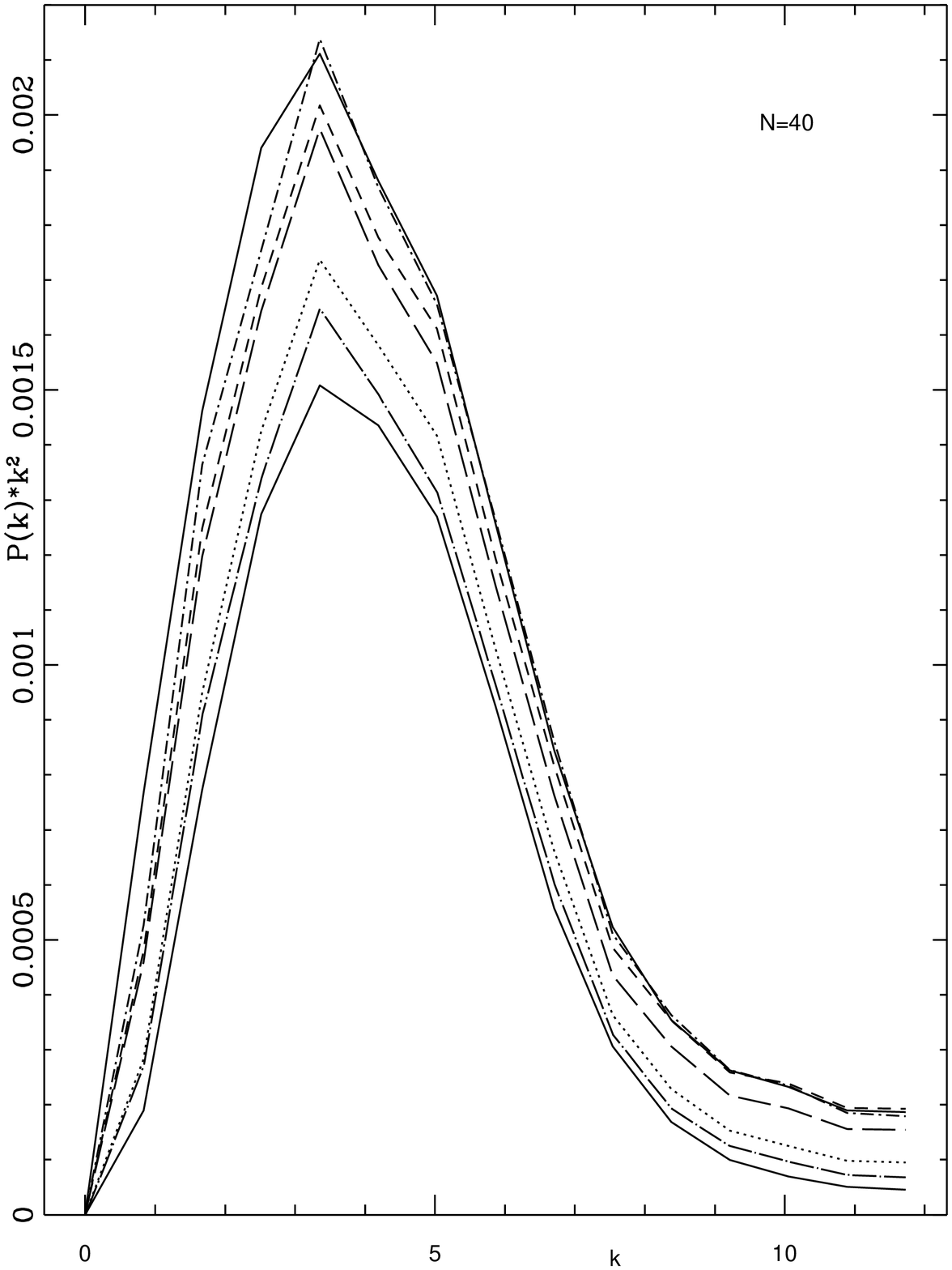 } {13}
To compare our results with those of Squires \& Kaiser (1996) we also
consider the case with no shear and surface density in the data
field (i.e.\ts the `noise-only-case').
This approach investigates the quality of the reconstruction (the 
quality of the `no-mass-detection')
for the case that there is no mass in the field at all, whereas we
have investigated before {\it how good} a two-dimensional 
mass distribution can be recovered. Given that a method which involves
a lot of smoothing will always fare better in the no-lens case than one
which spatially resolves noise, it is clear that the no-lens
comparison is not the relevant test [the ``best'' inversion in that
case is obtained by setting $\vc H\equiv \vc 0$ in (6)!].
%We point out that  it is then
%somewhat misleading to investigate a reconstruction method (i.e.\ts the
%KS-method)  which is not exact on a finite field. Because, e.g. \ts the
%$\D \equiv 0$ reconstruction kernel is also not exact, but 
%yields the by far best noise properties if there is no mass in the
%field. 

The curves in Fig.\ts3 demonstrate again that the noise properties of
NF and NF2 reconstructions are almost identical, whereas that of NF1
is different for reasons already discussed. For a dense grid ($N=80$)
all noise filtering methods become more and more equal, and the short
wavelength behavior approaches that of KS (solid line).  In any case,
the KS method is by far the best as long as there is no mass in the
field. As we already pointed out in Paper~I, this is because more (and
exact) information is used, namely that the shear is (set equal to)
zero outside the data field. The fact that the noise of the KS
inversion in Fig.\ts6 of Squires \& Kaiser (1996) is slightly larger
than that of the NF inversion at small wavelengths is due to the fact
that in their 
implementation of the KS algorithm, the shear field was not obtained
by smoothing the galaxy ellipticities, but the inversion was performed
by straight summation, which leads to shot noise (Seitz \& Schneider
1995).

Squires \& Kaiser (1996) suspected that the increase of noise of the
finite field inversion comes from the fact that they are more
sensitive to noise at the boundary of the data field.  This point is
clarified in Fig.\ts 4. The upper and lower solid curves denote the
power spectra for the NF and KS method on a $40\times 40$ grid. The
underlying galaxy distribution and thus shear field for each of the
individual reconstructions is by construction absolutely the same for
the NF and KS-case. We then embed the true data field $\cal U$ in a
two times as large field and distribute additional galaxies with the
same density and ellipticity distribution in the outer region. The
galaxies inside $\U$ are unchanged.  The shear field is calculated in
the large field and KS-reconstructions are obtained in the same
region. We cut out the surface mass density in the original field $\U$
and calculate the power spectrum of the reconstruction error in the
same way as for the other mass reconstructions within $\U$.  We point
out that in this case the shear field within $\U$ is not the same as
in the above case because now galaxies outside the field contribute to
the estimate of the shear field within $\U$. This makes the shear
field statistically smoother inside $\U$. But as can be seen in
Fig.\ts 4 (long-dashed-dotted curve) the reconstruction error within
$\U$ is larger than for KS-reconstructions of the small field (solid
line) -- because the data {\it outside} $\U$ are no longer `ideal
assumptions' ($\gamma\equiv 0$) but noisy measurements affected by the
intrinsic ellipticity distribution of the galaxies.  We then perform
reconstructions on the large field where the shear field is obtained
in the same way as before, but values on gridpoints within $\U$ are
substituted by the estimate obtained in the small field only. Thus the
$g$ field and its noise properties within $\U$ are now identical to
that of the KS and NF reconstructions of the small field. At the same
time the transition to the shear field outside $\U$ becomes less
continuous which mimics an artifical increase of noise at the boundary
of $\U$.  The power spectrum obtained from KS-reconstructions of that
$g$-field (dotted curve) is higher than the long-dashed-dotted curve,
as expected.

To obtain a KS-reconstruction where almost no information on data
outside $\U$ is used, we increase the noise outside $\U$ by doubling
the width of the ellipticity distribution for galaxies outside
$\U$. The shear field is calculated in the large field and the surface
density is KS-reconstructed. The power spectrum of the reconstruction
error within the small field is shown as short-dashed-dotted line --
and it is very similar to the power spectrum of the finite field
NF-reconstruction. One could argue that this large increase is caused
mainly by the fact that by the averaging procedure (8) the increased
noise outside $\U$ is partly tranferred in $\U$. To show that this is
not true we again consider the case where the shear field is
calculated in the large region as before, but where the values inside
$\U$ are the same as used for the KS- and NF-reconstruction of the
small field $\U$. We find that the reconstruction error is then only
marginally decreased (short-dashed curve). But still one could argue
that in this case the possibly non-smooth transition from the
$g$-field inside $\U$ to that outside could significantly contribute
to the noise. Therefore we smoothed that transition on the neighboring
gridpoints outside the data field.  The corresponding power spectrum
(long-dashed curve) shows that the smootheness of this transition has
only a small effect on the noise properties of the reconstruction
within $\U$. This comparison demonstrates that the KS-reconstruction
becomes worse the noisier the data outside $\U$ are and that the
assumption of $\gamma\equiv 0$ outside $\U$ is responsible for the
high quality of the KS-reconstruction {\it if} there is no mass in the
field.  Since this is not the case for most fields currently observed,
one is urged to use a method which is exact on finite fields (see
Squires \& Kaiser 1996).

Finally we apply the new noise filtering to the WFPC-2 geometry. 
Instead of performing a power-spectrum analysis, we have calculated
the mean-square deviation of the reconstructed density field $K(\vc
\theta)$ (shifted such that the mean value of $K$ over the field $\U$
equals the true one) from the input distribution (see also Fig.\ts 10
in Paper~I). We consider again two cases, the `no-lens-case' and 
 that of a mass distribution
which was now chosen similar to that in the Cluster Cl 0939 (see Fig. 5)
\xfigure{5}{This mass distribution  was used when the
rms-error for the NF and NF1 were compared; it was chosen to similar to
that of the cluster Cl0939. As in Fig. 1 the contours and surface plot
shows $- K(\vc x) = - \ln \eck{1
-\kappa (\vc x)} $. The grid is $40\times 40$ and the field of view is
$2.5$ arcminutes on a side.}
{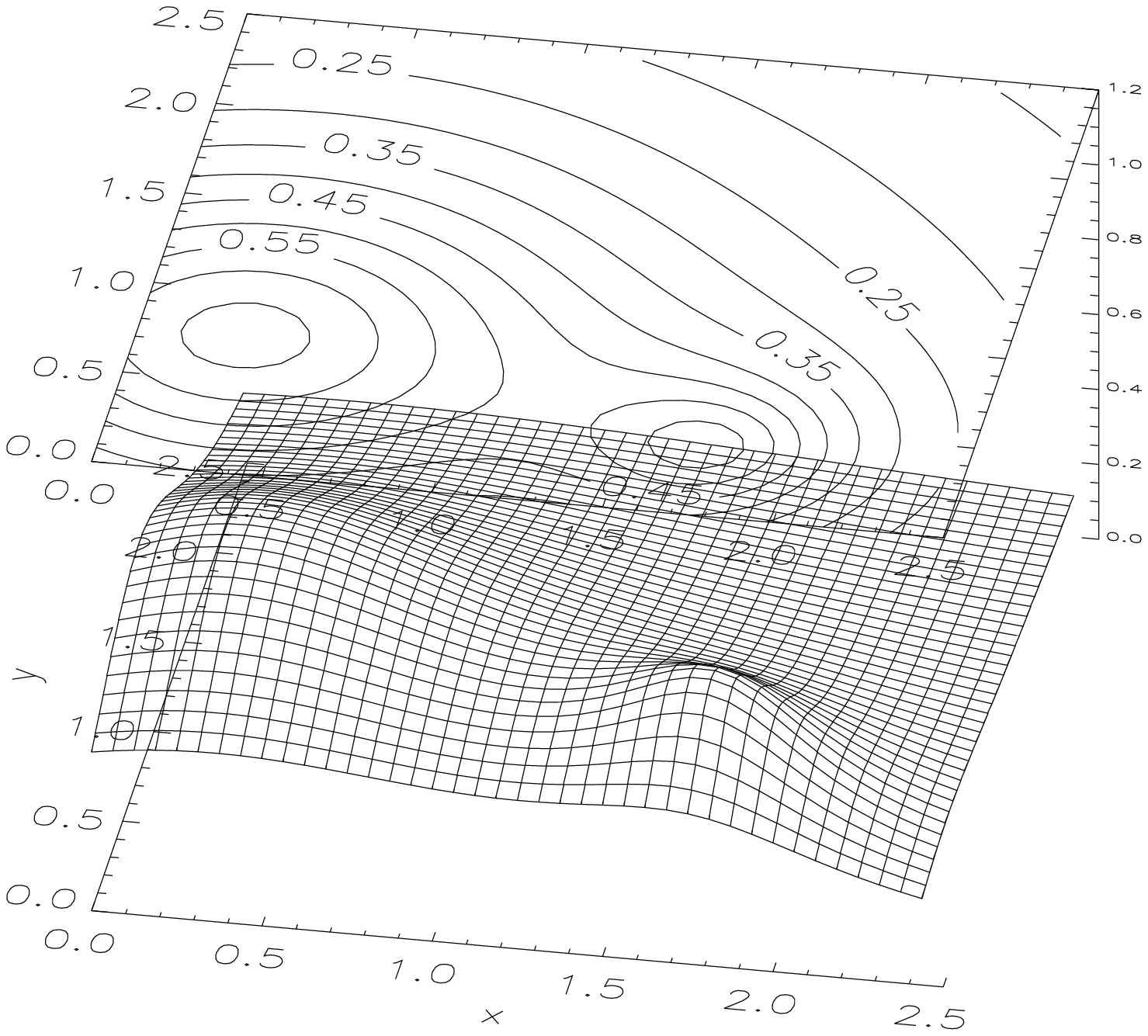 } {8}
 For both cases the galaxy density (60 per square
arcminute), the width of the ellipticity distribution ($\rho=0.2$) and
the smoothing length ($\Delta \theta =0\farcm 3$) were chosen equal to
the values for the weak lensing reconstruction of the cluster Cl0939
(Seitz et al. 1996). The reconstructions were obtained on a $40\times
40$ grid. 
In each of the two cases, two
different reconstructions are analyzed, one where the reconstruction
was performed on a square with $2\farcm 5$ sidelength, and the other
where one quarter of the square was removed. Fig.\ts 6 shows the rms
deviation for these cases, obtained from  500
 realizations for each case. For illustration,
only the WFPC-2 part of the square is shown in the first case. When
compared to reconstructions on the square, the WFPC-2 reconstructions
are just slightly more noisy close to the additional boundaries of the
field, owing to the smaller number of galaxies from which the shear is
obtained there. Note that the increase of noise at the `inner corner'
of the WFPC-2 is much smaller than that at the `outer corners', which
is due to the fact that at the former, more galaxies fall into the
filter scale than in the latter case. The increase of the noise at
those positions where the mass distribution peaks is due to the lack
of spatial resolution of the inversion, due to the smoothing
applied. In contrast to Paper~I we have not attempted here to adopt an
adaptive smoothing, depending on the lens signal, which would yield
better resolution near the mass peaks.
\xfigure{6}{...}
{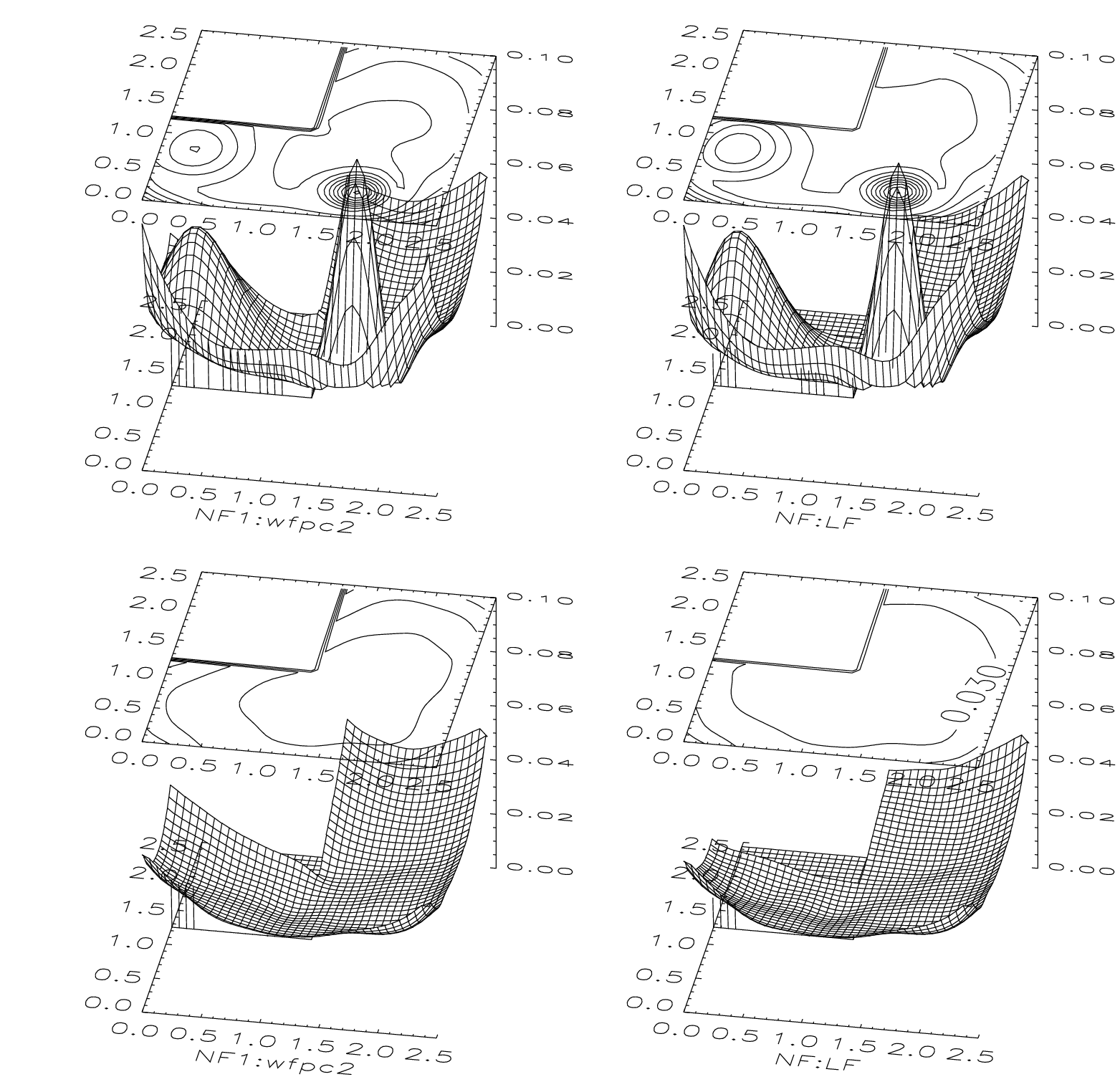} {14}

\sec{5 Conclusions}
We have derived a new version of the noise-filtering cluster mass
reconstruction algorithm originally proposed in Paper~I, which is
easier to implement, easier to use on large fields where the required
number of gridpoints can quickly exceed the number possible in using
the method of Paper~I,  
and which can easily be generalized to more
complicated geometries; the particularly relevant case of the WFPC-2
geometry was considered explicitly.
From extensive numerical tests we have shown
that the noise properties of this version is basically identical to
that of the method described in Paper~I. In agreement with Fig.\ts 6
of Squires \& Kaiser (1996), we conclude that the noise-filtering
method is the best known {\it direct} finite-field inversion
method. The comparison between the maximum probability method (Squires
\& Kaiser 1996) and the method presented here, carried out on the
mosaic of WFPC-2 centered on the cluster MS1358+62 (Hoekstra et al.\
1998), yielded no easily visible difference in performance of these
two methods.

We thank Bill Press for a very fruitful discussion which
triggered the work presented here. \SFB

\sec{Referenzen}
\def\ref#1{\vskip5pt\noindent\hangindent=40pt\hangafter=1 {#1}\par}
%\ref{Bartelmann, M. \ 1995, A\&A, 298, 661. }

\ref{Bartelmann, M. \ 1995, A\&A, 303, 643. }

%\ref{Bartelmann, M. \& Narayan, R. \ 1995, APJ, 451, 60.} 

\ref{Bartelmann, M., Narayan, R., Seitz, S. \&   Schneider, P. \ 1996,
APJL, 464, 115.}

%\ref{Blandford, R.D. \& Narayan, R.\  1986, APJ, 310 568.}

\ref{Broadhurst, T.J., Taylor, A.N. \& Peacock, J.A. \ 1995, ApJ, 438,
49.}

%\ref{Fahlman, G.G., Kaiser, N., Squires, G. \& Woods, D.  \  
% 1994, ApJ, 437, 56.}

\ref{Gorenstein, M.V., Falco, E.E. \& Shapiro, I.I. \  1988, APJ, 327, 693.}

\ref{Hoekstra, H., Franx, M., Kuijken, K. \& Squires, G.\ 1998, MNRAS,
submitted, also astro-ph/9711096.}

\ref{Kaiser, N. \& Squires, G.\ 1993, ApJ, 404, 441 (KS).}

\ref{Kaiser, N., Squires, G., Fahlmann, G.G., Woods, D. \&  Broadhurst,
T. \  1994, preprint astro-ph/9411029} %  (KSFWB).}

\ref{Kaiser, N.\   1995, ApJ, 493, L1.}

\ref{Kaiser, N., Squires, G. \&  Broadhurst, T. \  1995, ApJ, 449, 460.}

\ref{Lombardi, M., Bertin, G., 1998, astro-ph/9801244.}

\ref{Press, W.H., Teukolsky, S.A., Vettering, W.T. \& Flannery,
B.P. 1992, {\it Numerical Recipes}, Cambridge University Press.}

%\ref{Schneider, P., Ehlers, J. \& Falco, E.E. \  1992, {\it Gravitational
%Lenses},
%Springer-Verlag.}

\ref{Schneider, P. \  1995, A\&A,  302, 639.}

\ref{Schneider, P. \& Seitz, C.\ 1995, A\&A, 294, 411.}
%(Paper\ts I) }

\ref{Seitz, C. \& Schneider, P. \ 1995 A\&A, 297, 287.}% (Paper\ts
						       % II)}
\ref{Seitz, C., Kneib, J.P., Schneider, P. \& Seitz, S. \ 1996 A\&A,
314, 707.}

\ref{Seitz, C. \& Schneider, P. \  1997, A\&A, 318, 617.}

\ref{Seitz, S. \&  Schneider, P. \  1996,
%`Cluster lens reconstruction using only observed local data -- 
%an improved finite-field inversion technique', 
A\&A, 305, 383. (Paper~I)}

\ref{Squires, G., Kaiser, N., 1996, APJ, 473, 65.}

\ref{Tyson, J.A., Valdes, F. \& Wenk, R.A. \  1990, ApJ, 349, L1.}

%\endref{References}

\vfill\eject\end